\documentclass[pdflatex,sn-mathphys-num]{sn-jnl}

\usepackage{graphicx}%
\usepackage{multirow}%
\usepackage{amsmath,amssymb,amsfonts}%
\usepackage{amsthm}%
\usepackage{mathrsfs}%
\usepackage[title]{appendix}%
\usepackage{xcolor}%
\usepackage{textcomp}%
\usepackage{manyfoot}%
\usepackage{booktabs}%
\usepackage{algorithm}%
\usepackage{algorithmicx}%
\usepackage{algpseudocode}%
\usepackage{listings}%

\definecolor{magenta}{rgb}{0.965, 0, 0.859}

\theoremstyle{thmstyleone}%
\theoremstyle{thmstyletwo}%

\theoremstyle{thmstylethree}%

\begin{document}

\title[The Gerontocratization of Science: How hypergrowth reshapes knowledge circulation]{The Gerontocratization of Science: How hypergrowth reshapes knowledge circulation}


\author[1]{\fnm{Antoine} \sur{Houssard}}\email{antoine.houssard@cnrs.fr}
\equalcont{These authors contributed equally to this work.}

\author*[2]{\fnm{Floriana} \sur{Gargiulo}}\email{floriana.gargiulo@cnrs.fr}
\equalcont{These authors contributed equally to this work.}

\author[3]{\fnm{Tommaso} \sur{Venturini}}\email{tommaso.venturini@cnrs.fr}

\author[4]{\fnm{Paola} \sur{Tubaro}}\email{paola.tubaro@cnrs.fr}

\author[2,5]{\fnm{Gabriele} \sur{Di Bona}}\email{gabriele.dibona.work@gmail.com}
\equalcont{These authors contributed equally to this work.}

\affil[1]{\orgdiv{CIS}, \orgname{CNRS}, \orgaddress{\street{59 rue Pouchet}, \city{Paris}, \postcode{75017}, \country{France}}}

\affil[2]{\orgdiv{GEMASS}, \orgname{CNRS}, \orgaddress{\street{59 rue Pouchet}, \city{Paris}, \postcode{75017}, \country{France}}}

\affil[3]{\orgdiv{Medialab}, \orgname{Université  de Geneve}, \orgaddress{\street{ 24 rue du Général-Dufour}, \city{Geneve}, \postcode{1211}, \country{Switzerland}}}

\affil[4]{\orgdiv{CREST}, \orgname{ENSAE}, \orgaddress{\street{Institut Polytechnique de Paris}, \city{Palaiseau}, \postcode{91120}, \country{France}}}

\affil[5]{\orgdiv{Sony Computer Science Laboratories Rome, Joint Initiative CREF-Sony, Centro Ricerche Enrico Fermi}, \orgaddress{\street{Via Panisperna 89/A}, \city{Rome}, \postcode{I-00184}, \country{Italy}}
}


\abstract{%
Scientific literature has been growing exponentially for decades, with publications from the last twenty years now comprising 60\% of all academic output. 
While the impact of information overload on news and social-media consumption is well-documented, its consequences on scientific progress remain understudied.
Here, we investigate how this rapid expansion affects the circulation and exploitation of scientific ideas. 
Unlike other cultural domains, science is experiencing a decline in the proportion of highly influential papers and a slower turnover in its canons.
This results in the disproportionate persistence of established works, a phenomenon we term the ``gerontocratization of science''.
To test whether hypergrowth drives this trend, we develop a generative citation model that incorporates random discovery, cumulative advantage, and exponential growth of the scientific literature.
Our findings reveal that as scientific output expands exponentially, gerontocratization emerges and intensifies, reducing the influence of new research. Recognizing and understanding this mechanism is crucial for developing targeted strategies to sustain intellectual dynamism and ensure a balanced and healthy renewal of scientific knowledge.
}

\keywords{Scientific hypergrowth, Science of science, Information overload, Attention cycles}



\maketitle

\section{Introduction}
\label{sec1}

The scientific literature continues to expand at an exponential rate, with more than 60\% of all published research having appeared since the year 2000~\cite{priem2022openalexfullyopenindexscholarly}. In the last few decades, this exponential growth of scientific production has also been marked by an increase in the total number of people pursuing academic careers and in the average number of authors per paper~\cite{Ioannidis2018}.

Concerns about the overwhelming expansion of knowledge are not new. Even in the early 1600s, Barnaby Rich~\cite{Price1963-PRILSB-3} commented on the rapid proliferation of books, noting that:
\begin{quote}
    \textit{``One of the diseases of this age is the multiplicity of books; they doth so overcharge the world that it is not able to digest the abundance of idle matter that is every day hatched and brought forth into the world''.}
\end{quote}
To combat this ``disease'', scientific societies and journals were eventually established to manage and organize scholarly output. Ironically, however, these very institutions evolved into key drivers of the explosive growth in academic publications that they originally sought to regulate.
In the 1960s, Derek De Solla Price highlighted the risk that \textit{science hypergrowth} could overwhelm researchers, making it difficult for them to keep up with new developments and to draw attention to their new work~\cite{Price1963-PRILSB-3,Bornmann2021,Larivire2007}. Nonetheless, following Malthusian ideas~\cite{malthus1986essay}, he also predicted that the initial exponential expansion would eventually give way to a saturation or stationary phase. However, the combined effect of the proliferation of research and higher-education investments, the development of technologies that facilitate the writing and dissemination of papers~\cite{https://doi.org/10.48550/arxiv.1411.0275}, the rise of a market for academic publishers (as well as of ``mega-'' and ``predatory journals''~\cite{Spezi2017}), and the consolidation of a ``publish or perish'' ethos~\cite{Sarewitz2016PublicationFlood} ultimately proved him wrong.

Although the phenomenon of scientific hypergrowth is well-recognized, its specific impacts on the structure and dynamics of scientific ecosystems remain underexplored. While extensively discussed by classic authors in the sociology of science~\cite{merton1968matthew,Collins1977-ej}, this issue has not been fully examined in its current and concrete implications on the structure and dynamics of the scientific publication system.

\medskip

The hypergrowth dynamic of the scientific ecosystem resembles, to some extent, the acceleration of information flows that has been observed in other cultural sectors with the advent of digital media~\cite{LorenzSpreen2019,Candia2018}, a defining characteristic of postmodern societies~\cite{Rosa2010-pz}. In media studies, the impact of information overload has been widely discussed in relation to the affordances of social media and the effects of their recommendation algorithms. Information overload has also been linked to different phenomena including political polarization, fake news diffusion, and the formation of echo chambers and filter bubbles.

The explosion of user-generated content on online platforms, combined with a growing reliance on metrics such as clicks and views~\cite{Bergstrm2018,costera2015checking,kormelink2018clicks}, has been connected to the acceleration of news cycles~\cite{LorenzSpreen2019,Candia2018} and to a superficial mode of consumption characterized by a quick turnover in interest and a renewed relevance of ``opinion'' leaders~\cite{Karlsen2015,Bergstrm2018}. The attention regime typical of online social networks has been in fact described as dominated by ``junk news bubbles'', i.e., sudden bursts of interest around pieces of news or matters of discussion that capture large shares of collective attention, but only for a very short time. 
These bubbles often focus on sensational, ephemeral topics that are entertaining but lack substantial informational quality, hence the label ``junk news''.
It has been shown that attention bubbles can be generated by the influence of recommendation algorithms and their preference for trendy contents~\cite{castaldo2022junk}, and that this phenomenon is amplified when the information flow is larger.

\medskip


While the popularity of online content is measured through views and comments, the popularity of a scientific publication is generally measured by the number of citations it attracts. Even though this measure is not exempt from criticism~\cite{Bornmann2008}, citations allow gauging the intensity with which arguments, ideas, articles, people, research programs, disciplines, etc.\ resonate in the scientific community~\cite{Klamer2002}. 
Research on the evolution of citations has produced diverging results. While some studies have suggested a lengthening in citation life cycle~\cite{https://doi.org/10.48550/arxiv.1411.0275,Bouabid2013,Wallace2012}, others have shown constant or accelerating attention cycles~\cite{LorenzSpreen2019}; some have highlighted an increase in citation inequality~\cite{Barabsi2012}, others a decrease~\cite{Pan2018}; some have observed the persistence of key concepts~\cite{chu2021slowed} even after they stop being directly referenced~\cite{meng2023hidden}, and others the decreasing disruptiveness of new publications~\cite{petersen2024disruptionindexsufferscitation}.

None of these studies, however, has specifically investigated the link between science hypergrowth and attention disorders in the scientific ecosystem. To zoom in on this question, here we analyze the citation patterns in various disciplines, using datasets extracted from OpenAlex, an open-access database of scientific publications. In particular, we note a pattern that is distinctively different from that of news and digital media: while we observe an increasing concentration of academic attention on a minority of publications, such papers become more and more persistent in the popularity ranking. This low turnover rate of the most popular publications suggests a phenomenon of ``standardization of scientific canons'' and ``aging of the scientific elites''.





Scientific hypergrowth, however, is not the only possible culprit for this ``gerontocratization''. Other factors, such as the platformization of science and the consequent action of recommendation algorithms, the advent of predatory journals (and the different trust regime that they creates) and individual behaviors in the research community, may also contribute to the sclerotization of the cumulative advantage process. 
To disentangle these concurrent mechanisms, we employ simulations designed to test whether scientific hypergrowth is directly responsible for gerontocratization. Our generative model, introduced and analyzed in section \ref{Q2}, shows that, excluding all other factors, hypergrowth can lead to stagnation within the scientific ecosystem. 



\section{Results}
\subsection{Rising average impact despite science hypergrowth}
\label{Q1}

\begin{figure}[tb]
    \centering
    \includegraphics[width=\linewidth]{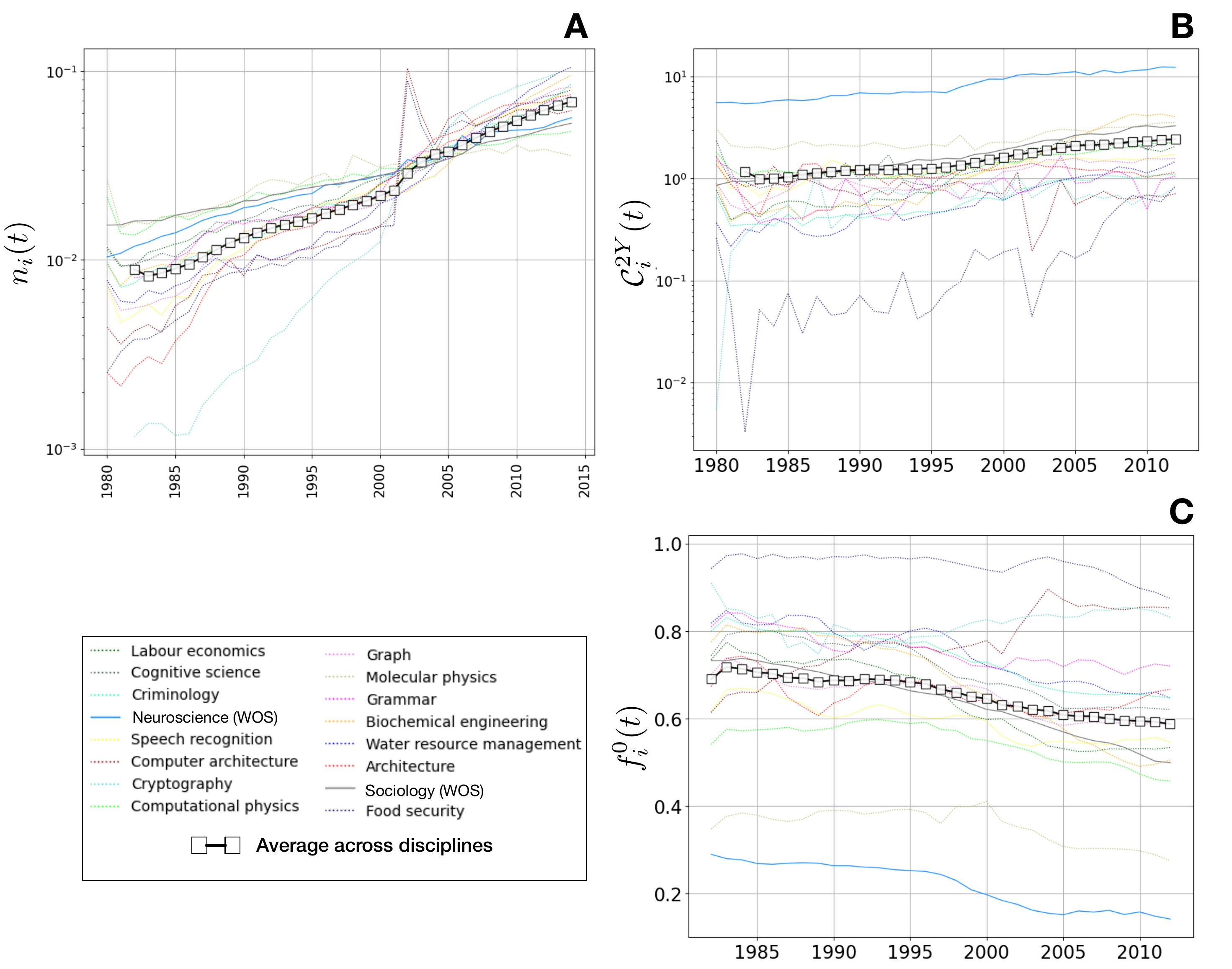}
    \caption{\textbf{(A)} Relative number of publications by year for all the disciplines. \textbf{(B)} Average number of citations received by papers appeared in year $t$, after 2 years from their publication. \textbf{(C)} Fraction of papers, appeared in year $t$, without any citation after two years from their publications. 
    \\
    Each discipline is characterized by a different color, while the square points represent the average measures among all datasets.}
    \label{fig1}
\end{figure}

Following the results of De Solla Price~\cite{Price1963-PRILSB-3}, we first quantitatively analyze the relative growth of the number of papers in time for various disciplines (see section~\ref{data} and~\ref{data_collection} for details on the data sources and data collection process). In Fig.~\ref{fig1}A, we plot the normalized number of papers $n_i(t)$ published every year in each discipline $i$, that is
\begin{equation}
n_i(t)=\frac{N_i(t)}{\sum_t N_i(t)},
\end{equation}
where $N_i(t)$ is the number of papers published in discipline $i$ in year $t$. 

As already observed in previous works~\cite{priem2022openalexfullyopenindexscholarly, Bornmann2021,Larivire2007}, the normalized number of publications increases in time with an exponential growth regime, i.e., we can fit such growth with the expression 
\begin{equation}
    n(t)=n_0e^{\alpha t},
    \label{eq:gr}
\end{equation}
where the exponents and coefficients depend on the specific field of study. The different disciplines we examine exhibit the same structural behavior, with exponent $\alpha$ varying between $0.004$ for Molecular Physics to $0.14$ for Biochemical Engineering.

Averaging across all disciplines, we find that 91\% of papers receive their first citation in the first two years of publication, with limited variations between each discipline. 
To measure the initial attraction of citations of each paper $x$, let us first indicate with $t_0(x)$ its publication year.
Let us also indicate with $k_x^i(t)$ the number of citations that a paper $x$, belonging to discipline $i$, receives in a year $t$.
Then, we count the average number of citations that the papers in discipline $i$, published in a year $t$, received in the first two years after publication ($2Y = [t,t+2]$), that is $C_i^{2Y}(t)=\langle \sum_{t'=t}^{t+2} k_x^i(t')\rangle_{x\in i|t_0(x)=t}$.
In Fig.~\ref{fig1}B, we can observe that, as the year of publication moves forward, papers receive more and more citations within the first two years after publication. Similarly, in Fig.~\ref{fig1}C we can see that the portion $f_0^{2Y}(t)$ of ``invisible literature'', namely the fraction of papers published in year $t$ that do not receive any citations in the first two years of life, decreases linearly in time for all the disciplines.

We could conclude from these first observations that, although the number of papers published each year is exponentially growing, scientific publications have gained over time a larger capacity to circulate and to be recognized, through citations, in the academic ecosystem soon after being published.

\subsection{Science is not getting ``junk''}

Given the exponential growth of scientific literature, we extend our previous analysis to investigate whether the initial surge of attention some publications receive may signal the emergence of ``junk science bubbles''. 
These bubbles refer to ephemeral scientific works that would attract a large share of attention in a short period of time, with a rapid turnover of the most popular ones, similar to trends observed for other cultural products, such as videos, songs, etc.~\cite{castaldo2022junk,LorenzSpreen2019}.

Let us define the attention on a paper $x$ utilizing the number of citations $k_x^i(t)$ received by $x$ from papers of discipline $i$ published in year $t$.
To remove the bias due to the exponential expansion of the scientific ecosystem, for each paper $x$ we define the attention share $\xi_x^i(t)$ of a paper $x$ for each year $t$ by normalizing the number of citations $k_x^i(t)$ as follows:
\begin{equation}
    \xi_x^i(t)=k_x^i(t)\Bigg/\sum_{y\in i} k_y^i(t).
\end{equation}

We begin by examining how attention is distributed among competing content within each disciplinary arena. Specifically, we analyze the evolution of the Gini index of the citation share $\xi_x^i(t)$ for each discipline on a year-by-year basis. 
Ranging from 0 (perfect equality) to 1 (extreme inequality), the Gini index is a widely used indicator in sociology and economics that measures the level of inequality for a specific variable within a given population~\cite{gini}. In the context of our study, it allows us to quantify the inequality in the distribution of citation shares among publications within each discipline. 

\begin{figure}[tb]
    \centering
    \includegraphics[width=\linewidth]{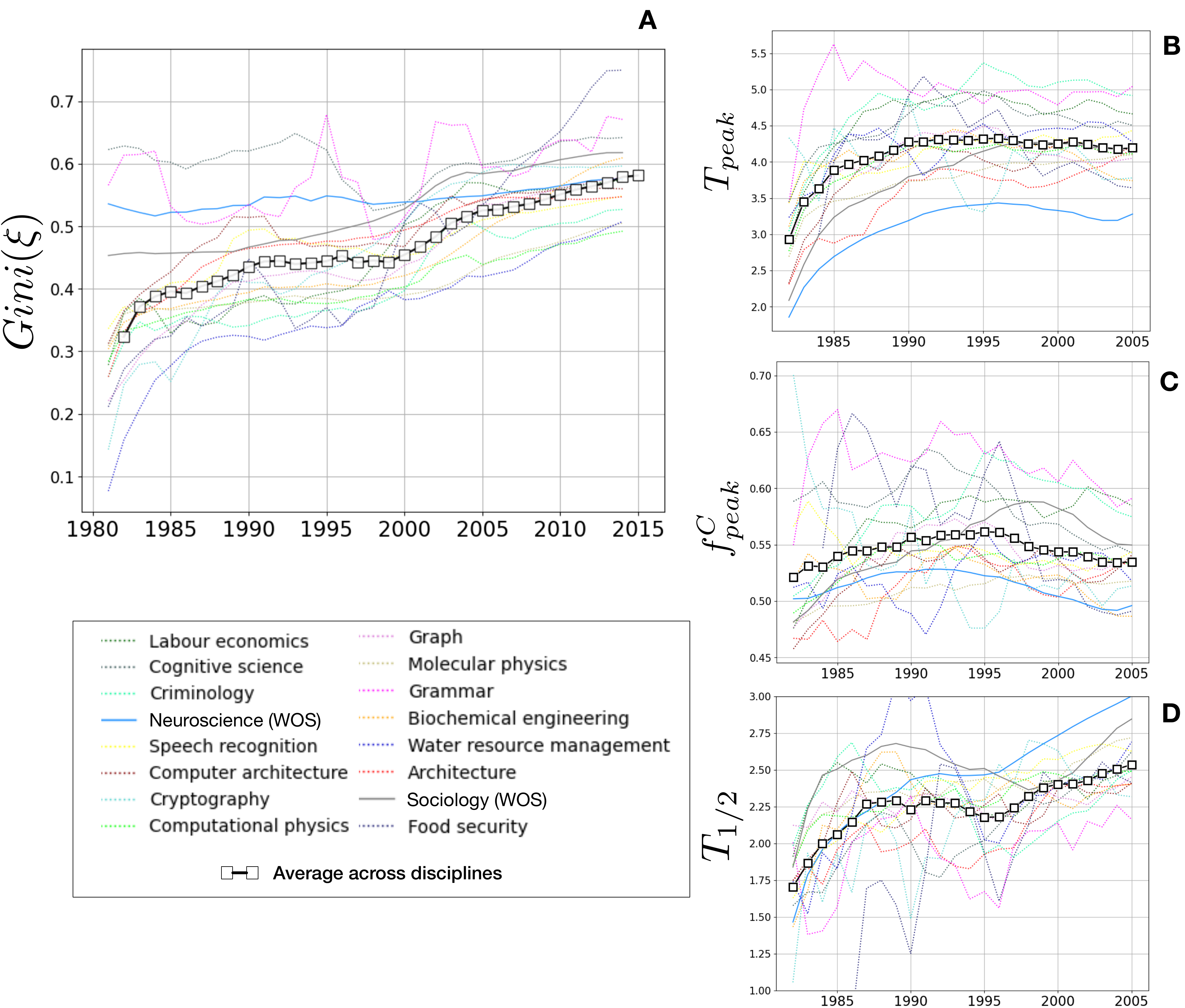}
    \caption{\textbf{(A)} Gini index of the citation share $\xi_x^i(t)$ for each discipline on a year-by-year basis. 
    \textbf{(B)} Evolution of the time $T_{peak}$ from the publication year to reach the peak with maximum share, averaged across papers published each year in each discipline.
    \textbf{(C)} Evolution of the cumulative fraction of share at the peak $f^C_{peak}$.
    \textbf{(D)} Evolution of the half-life time $T_{1/2}$ from the peak.\\
    Each discipline is characterized by a different color, while the square points represent the average measures among all datasets.}
    \label{fig2}
\end{figure}

In Fig.~\ref{fig2}A we show that the Gini index ($Gini(\xi)$) increases over time for all disciplines, indicating a more and more unequal distribution of the citation shares, similarly to what previously observed in different case studies~\cite{kozlowski2024decrease}.
This suggests that, much like other forms of cultural contents, scientific literature exhibits an increasingly unequal structure, with fewer papers able to monopolize the attention space. 

To test the junkisation hypothesis in science, we also need to analyze attention cycles, namely the shape of the attention share patterns related to each paper over time.
To compare papers appearing in different years, we consider the share curves limited to the first 10 years following publication. We normalize each share curve by dividing each point by the area under the curve, obtaining the attention share fraction $\tilde{\xi}_x^i(t)$ in each year $t$, that is:
\begin{equation}
    \tilde{\xi}_x^i(t)=\xi_x^i(t)\Bigg/\sum_{t_0(x)<t'<t_0(x)+10} \xi_x^i(t').
\end{equation}
where $t_0(x)$ is the publication year of paper $x$.
 
We characterize these patterns through three different measures~\cite{parolo2015attention}: (1) the time $T_{peak}$ from the publication year to reach the peak with maximum share, averaged across papers published each year in a discipline; (2) the cumulative fraction of share at such peak, defined as $f^C_{peak} = \sum_{t \leq T_{peak}}\tilde{\xi}_x^i(t)$; and (3) the half-life time $T_{1/2}$ from the peak, namely the time after the peak needed to reach the last point where the curve exhibits a value equal to the half of the maximum value of $\tilde{\xi}_x^i(t)$.
To reduce statistical noise in this part of the study, we only consider papers which have received at least ten citations in total. Moreover, to avoid getting incomplete cycles, we only focus on papers published before 2006.

We observe in Fig.~\ref{fig2}B that the average time $T_{peak}$ to reach the popularity peak, for all disciplines, increased by 1.5 years from the '80s to the 90's, stabilizing afterwards to a specific value typical of each discipline. The frequency $f^C_{peak}$ at the peak, instead, does not show a significant variation in time (Fig.~\ref{fig2}C). Finally, measuring the attention decay by the half-life time from the peak (Fig.~\ref{fig2}D), we observe that the average half-life $T_{1/2}$ of papers increases with time in all disciplines. The combination of these measures indicates that the citation cycles of papers exhibit a more durable persistence in the scientific ecosystem, i.e., they are not subject to ephemeral trends, on average.  

\medskip

In summary, while we have observed an increase in inequality patterns, we also notice a lengthening of the attention period. As we are about to see, this increased attention appears especially true for the most important papers, in contrast with the ``junkisation'' phenomenon observed in other types of online content.   

\subsection{Canons get more stable and ``elites'' get older}
The higher concentration of citations on a small set of papers, together with the longer persistence of attention on papers observed in the previous section, potentially impacts the structuring of the pillars of the literature. Indeed, the combination of these phenomena suggests that the ranking of such pillars becomes more and more stable. To confirm this hypothesis, we compute the similarity between the top-$k$ list of papers---the list of papers occupying the first $k$ positions in the ranking based on the number of citations received in a specific year---of two consecutive years using the ``ranked Jaccard similarity'', a measure to compare rankings previously introduced in Ref.~\cite{gargiulo2016classical}. 
From each top-$k$ list we build a new list, that we name expanded-top-$k$ list, where each element, at position $r$ of the top-$k$, is identically repeated $k-r$ times. 
Then, the Jaccard similarity is calculated between such expanded lists. 

This measure allows one to consider both the position in the ranking and possible changes in the top-$k$ lists from one year to another. 
On the one hand, it indeed takes into account the position in the ranking, unlike the traditional Jaccard similarity, which ignores ranking order. On the other hand, it also captures changes in the top-$k$ list, unlike other traditional methods for comparing top-$k$ lists, like the Kendall-Tau index.

\medskip

\begin{figure}[tb]
    \centering
    \includegraphics[width=\linewidth]{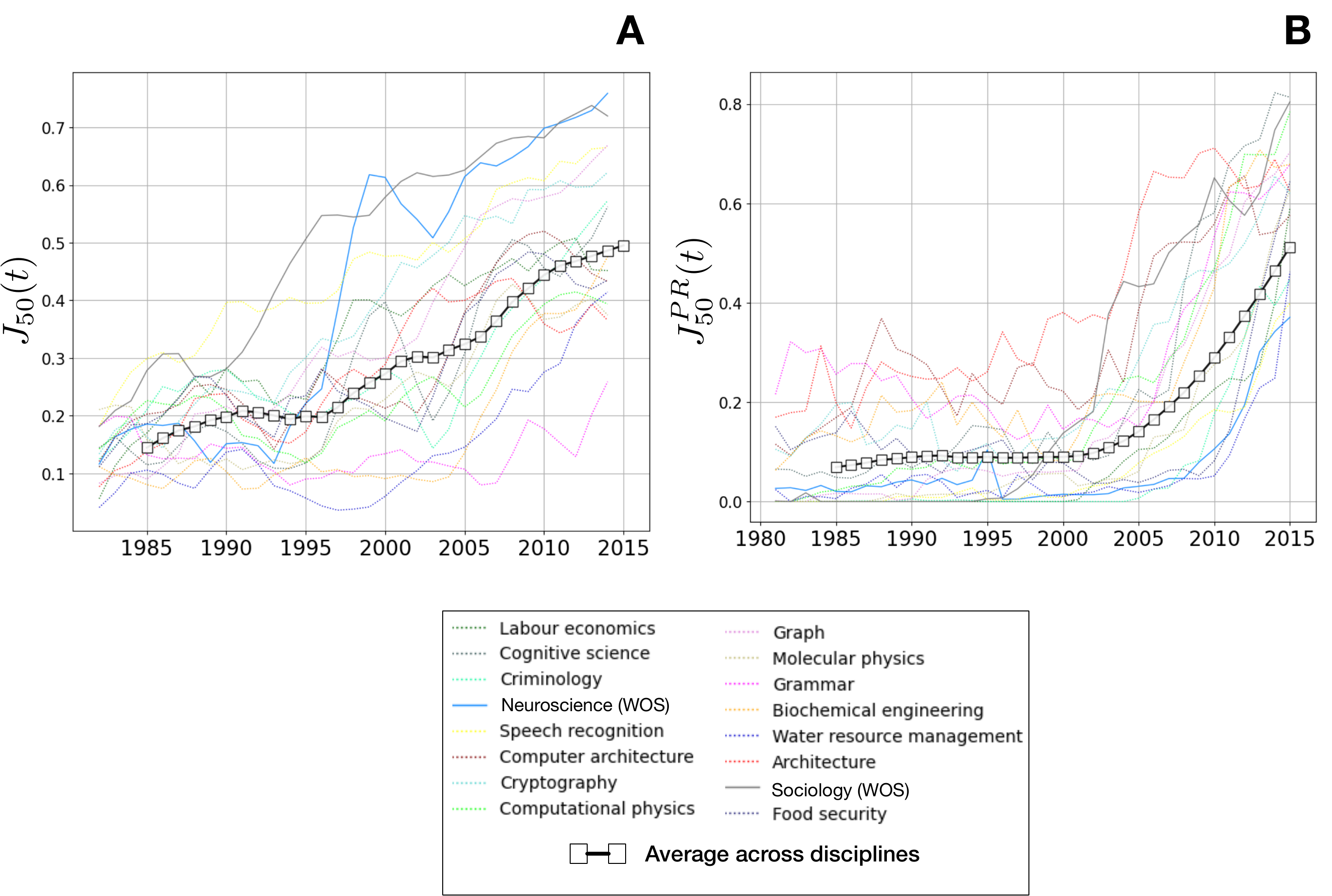}
    \caption{\textbf{(A)} Ranked Jaccard similarity $J_{50}(t)$ between the top 50 papers in the citation ranking at year $Y$ and the ones at $Y-1$. \textbf{(B)} Ranked Jaccard similarity $J_{50}^{PR}(t)$of the top 50 papers according to Page Rank centrality in the citation network at $Y$ and those at $Y-1$s.
    }
    \label{fig3}
\end{figure}

In  Fig.~\ref{fig3}A, we show the ranked Jaccard similarity between the top-$k$ lists at two subsequent years for all disciplines (and for $k=50$). Similar results are obtained for different values of $k$. Notice the significant increase of the ranked Jaccard similarity over time for all disciplines, meaning that the list of top-50 cited papers becomes more and more similar among consecutive years. 
The same results can be observed, in Fig.~\ref{fig3}B, if we consider the top-50 papers according to the PageRank measure in the temporal citation networks. 
Here, the citation network of year $t$ is defined as a directed graph in which the nodes represent papers published in year $t$, along with all other papers that cite or are cited by them. An edge between two nodes is drawn if one cites the other. 
Notably, we observe a significant increase since the first years of this century.

These results show that the scientific pillars are stabilizing, even though there is an exponential growth of literature, giving origin to a lack of renewal of the canons.

\medskip

\begin{figure}[tb]
    \centering
    \includegraphics[width=\linewidth]{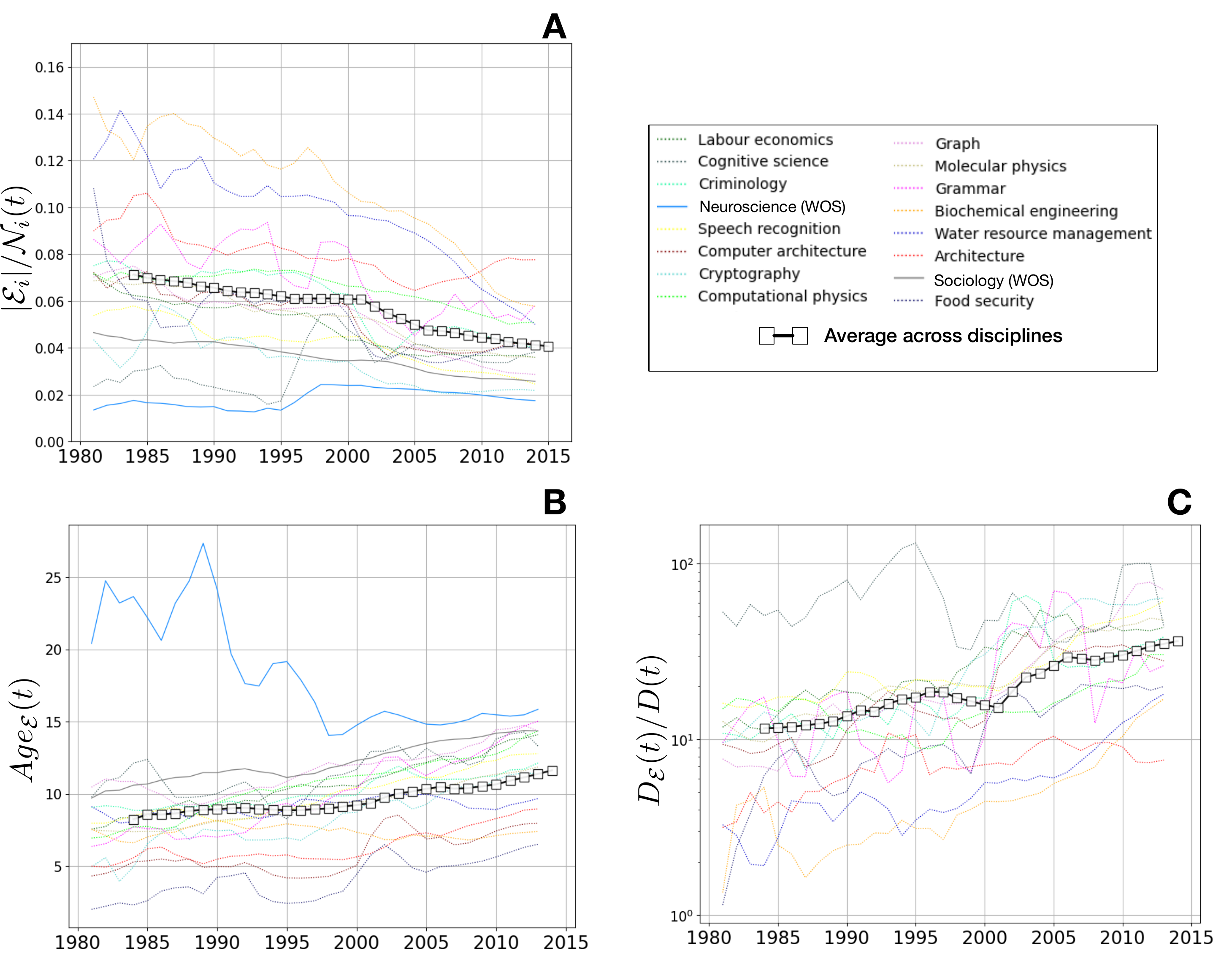}
    \caption{
    \textbf{(A)} Relative size of the set of elite papers in time. 
    \textbf{(B)} Average age of elite papers in time. 
    \textbf{(C)} Temporal evolution of the relative co-citation density in the sub-graph of elite papers with respect to the whole co-citation network.
    }
    \label{fig4}
\end{figure}

To better analyze the concentration phenomenon of citations in a restricted number of papers, we analyze the structure of the ``elite'' papers, $\mathcal{E}(t)$, defined as the set of papers that together get the 80\% of the total number citations in year $t$. While different between disciplines, the size of the elite set of papers, $|\mathcal{E}(t)|$, is relatively small and, above all, it displays a clear linear decrease in time (Fig.~\ref{fig4}A).

Moreover, approximating the age of a paper with the number of years from its publication, we define the age of the elite set at a certain year $t$ as the average age of the papers belonging to this set.
As shown in Fig.~\ref{fig4}B, the age of the elite set in all disciplines increased of almost five years during the observation period, with the exception of neuroscience that displays an initial noisy decrease until the year 2000, before starting to increase.

\medskip

Inspired by rich-club measures in network science~\cite{colizza2006detecting}, we finally investigate if there is an emerging tendency to cite, within the same paper, papers that are part of the elite, or if the elite papers are disconnected among them, each having its own influence on an isolated part of the scientific literature (i.e., scientific sub-communities are disconnected and each of them has its own important reference). 
For each year $t$, we analyzed the temporal co-citation network, where nodes represent the papers cited by other papers published in year $t$ and are linked if they appear together in the reference list of one of these papers. 
We hence compared the density of the sub-graph containing only the elite nodes with the density of the whole co-citation network (Fig.~\ref{fig4}C). We observe that this relative density increases over time, strengthening the connections among elite papers and reinforcing a progressively more pronounced core-periphery structure. In this structure, elite papers are not only over-cited but also frequently cited together.

\medskip

To summarize, the peculiar attention regime observed in the scientific ecosystem---characterized by increasing inequality in attention shares and longer cycles of attention on papers---is associated to a general consolidation of the canonical references and the emergence of an elite structure.
This elite becomes progressively smaller in relative size while its average age increases over time. Furthermore, this elite set of papers grows increasingly interconnected compared to the rest of the literature, meaning that elites play an increasingly central role in bridging the knowledge space.
Based on these observations, we can describe this as a ``gerontocratic'' reorganization of science.

\subsection{Origins of the scientific ``gerontocracy''}
The gerontocratic reorganization of science raises a fundamental question: what drives this phenomenon? Is it triggered by exogenous factors, such as the introduction of recommendation algorithms? Does it stem from shifts in scientific practices, including the ``publish or perish" culture or the proliferation of journals? Or is it an inherent consequence of the hypergrowth regime of the scientific ecosystem?

A first hint in the direction of the last hypothesis comes from the Heaps' law of citations. Heaps' laws~\cite{tria2018zipf} were originally introduced in linguistics to analyze the growth patterns of vocabularies. According to the Heaps' law, the number $D$ of different words encountered while reading a text grows as a power-law function of the total number of words $N$ used up to that point, i.e., $D(N) \sim N^\beta$, where the Heaps' exponent $\beta$ indicates the pace at which new words are used. A low value of $\beta$ indicates that, as the system grows, fewer new words are used and, more generally, fewer new discoveries are made. 

\begin{figure}[tb]
    \centering
    \includegraphics[width=\linewidth]{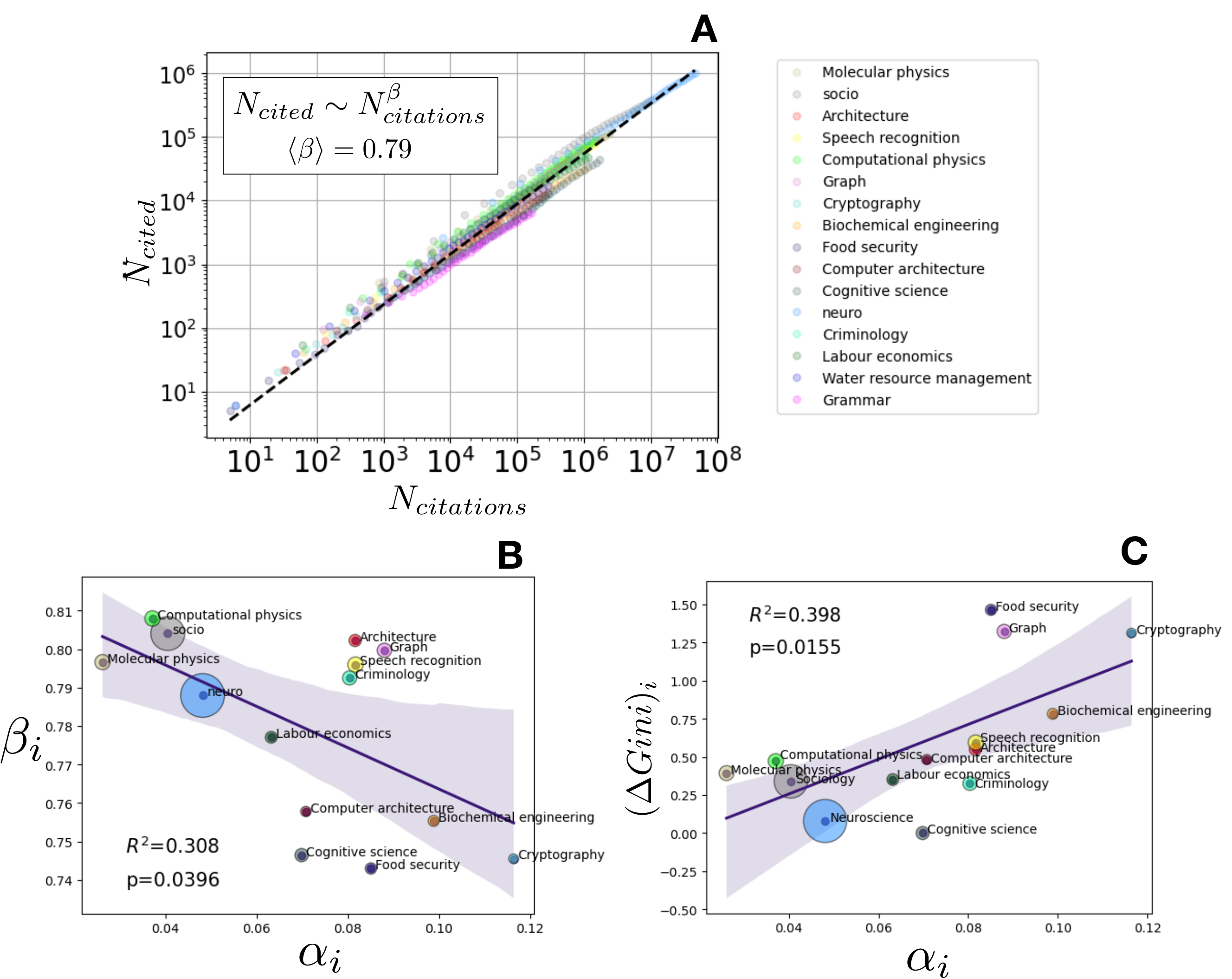}
    \caption{\textbf{(A)} Heaps' law for the different disciplines, showing the evolution of the number of different cited papers as a function of the total number of citations. 
    \textbf{(B)} Correlation between the exponent of the Heaps' law ($\beta$) and the growing exponent of the corpus size ($\alpha$).Each point is a discipline. The size of the points is proportional to the total size of the dataset. 
    \textbf{(C)} Correlation between the realtive variation (between 1980 and 2015) of the Gini index  and the growing exponent of the corpus size ($\alpha$). Each point is a discipline. The size of the points is proportional to the total size of the dataset. 
    }
    \label{fig5}
\end{figure}

In our case, considering the time-ordered aggregation of papers' reference lists as a large ``text'', where references act as words, we counted the total number of different cited papers, $N_{cited}$, with respect to the total number of citations $N_{citations}$ (see Fig.~\ref{fig5}A). We found that all disciplines follow a Heaps' law with a very similar Heaps' exponent $\beta$, varying between 0.73 and 0.81. Fitting all the disciplines together we obtain the global Heaps' law, $N_{cited}\sim N_{citations}^{0.79}$. 
The universality and the stationarity of this law suggest that the ``aging''  patterns, favoring the persistence of attention on a few older works, could be a direct consequence of a universal microscopic dynamic underlying the morphogenesis of citation networks, that is, their structural evolution~\cite{pastor2004evolution}. 
Moreover, the regression plot in Fig.~\ref{fig5}B shows that the Heaps' exponent $\beta$ for each discipline is negatively correlated with the system's growth rate $\alpha$, i.e., the exponent $\alpha$ of the growing curve for the fraction of papers in Eq.~\eqref{eq:gr}. This suggests a hidden link between growth and discovery processes. 
In Fig.~\ref{fig5}C, we further show that the variation of the Gini index for each discipline, defined as 
$\Delta Gini= (Gini(2015)-Gini(1980))/Gini(1980)$, is positively correlated to the growth rate $\alpha$ of the system.

These findings seem to suggest that, beyond other potential contributions, the process driving science to the increase of citations' concentration into a few old ``elite'' papers (gerontocratization) might have a direct causal connection with the exponential growth of the system. These patterns could stem from a microscopic process in which the fundamental morphogenetic rules remain largely constant over time, while the system itself expands exponentially.

\subsection{Hypergrowth drives gerontocratization}
\label{Q2}


To test the hypothesis proposed in the previous paragraph---that gerontocratic patterns can emerge directly as a consequence of hypergrowth---we need to isolate the effect of system growth from all the other factors that may influence the formation of the citation network. 
Since it is not possible to isolate these effects in empirical data, we construct a simple generative citation model in which the only relevant variable is the expansion rate of the literature.

Our model draws inspiration from Pólya’s urn model and the urn model with triggering~\cite{tria2014dynamics,tria2018zipf}, a well-established approach for modelling discovery processes. Our model simulates the reference selection process for new articles entering the system, assuming that the number of new papers grows according to an exponential expansion rule, $\mathcal{N}(t) = \mathcal{N}_0e^{\alpha t}$. Here, the parameter $\alpha$ controls the expansion rate of the system, with $\alpha=0$ representing linear growth.
The mechanism driving citation selection in our generative model is based on the interplay of two processes. On the one hand, citations can follow a cumulative advantage mechanism (the Matthew effect)~\cite{merton1968matthew}. 
On the other hand, citations can result from an exploration process, leading to the discovery of unknown or newly published literature.

\medskip

\begin{figure}[tb]
    \centering
    \includegraphics[width=\linewidth]{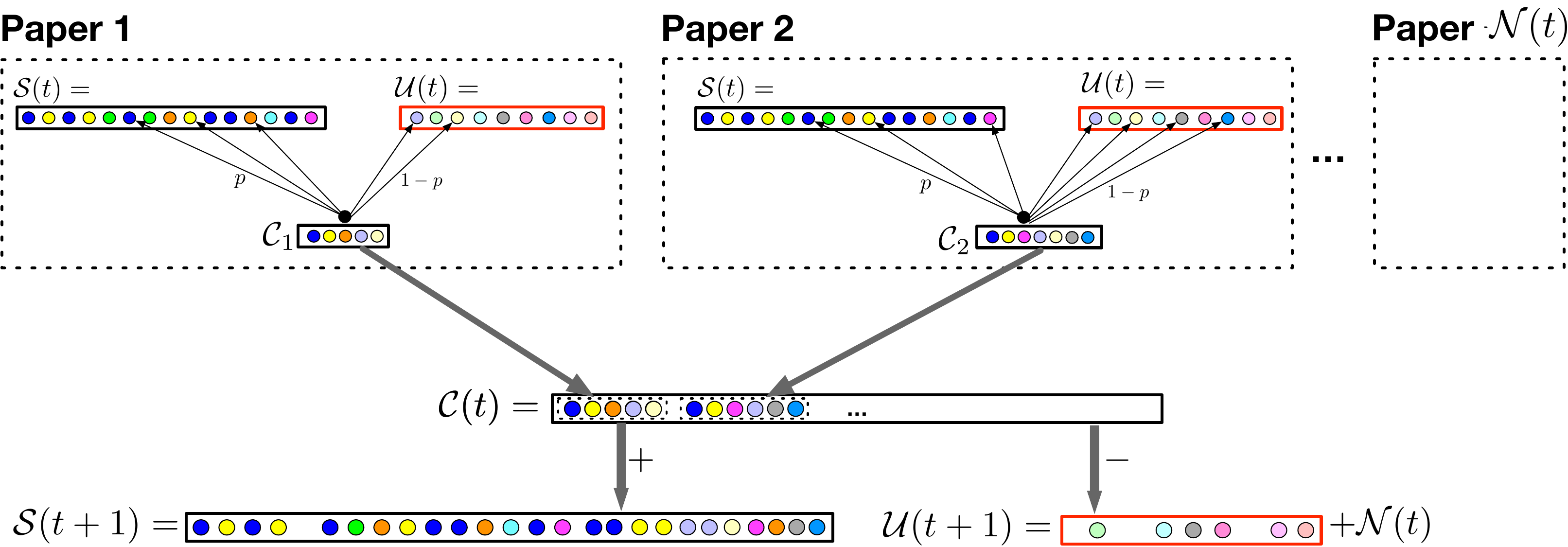}
    \caption{Schematic illustration of an iteration of the model. At time $t$, $\mathcal{N}(t)$ new papers enter the system. Each paper $i$ forms its reference list $C_i$, extracting papers from $\mathcal{S}(t)$ with probability $p$ and from $\mathcal{U}(t)$ with probability $1-p$. All reference lists are then combined together in $\mathcal{C}(t)$. Each reference from $\mathcal{C}(t)$ is added to $\mathcal{S}(t+1)$, while any reference extracted from $\mathcal{U}(t)$ or with an age older than 2 time steps is removed from $\mathcal{U}(t+1)$. Finally, all the new $\mathcal{N}(t)$ papers are added to $\mathcal{U}(t+1)$.}
    \label{fig6}
\end{figure}

The basic steps of the model are illustrated in Fig.~\ref{fig6}. At each time step $t$, corresponding to a year, $\mathcal{N}(t)$ new papers enter the system, with each of them selecting its references. 
For each of the $N_{ref}$ references in a paper's bibliography, 
the selection process follows two mechanisms: cumulative advantage and exploration. With probability $p$, a new paper cites an already cited work, reinforcing its prominence through a cumulative advantage mechanism (the Matthew effect). With probability $1-p$, the paper instead engages in an exploration process, selecting a previously uncited work.

To implement these mechanisms, we use two urn models. Previously cited papers are drawn from a first urn $\mathcal{S}$, where each paper appears as many times as its number of prior citations, ensuring preferential attachment to well-cited works. In contrast, uncited papers are selected uniformly from a second urn $\mathcal{U}$, which contains all uncited published papers.
Considering that approximately 90\% of papers receive their first citation within two years, papers older than two time steps are removed from $\mathcal{U}$, becoming effectively invisible literature. 

Summarizing, at each time step $t$, the new papers simultaneously build their bibliographies based on the contents of $\mathcal{S}(t)$ and $\mathcal{U}(t)$. Once all bibliographies are created, the cumulative list of citations for that year, $\mathcal{C}(t)$, is compiled. Papers in $\mathcal{C}(t)$ are then added to $\mathcal{S}$, including all repetitions, and removed from $\mathcal{U}$ if they were previously uncited. Finally, the $\mathcal{N}(t)$ new papers are added to $\mathcal{U}$, and papers older than two years are removed from it.
More details are provided in the methods section \ref{methods-model}.

\medskip

The model allows to analyze the direct effect of the system growth---represented by the growth parameter $\alpha$---on three key aspects observed in the data: the concentration of citations on canonical works, measured by the Gini index on the distribution of citations; the stabilization of elite publications, tracked by changes in the ranked Jaccard similarity of the most cited papers each year; and the discovery process of new literature, assessed through the Heaps' laws. 
For each value of the parameter $\alpha$, we hence compute these three measures on each replica of the simulation and analyze their averages.

\begin{figure}[tb]
    \centering
    \includegraphics[width=\linewidth]{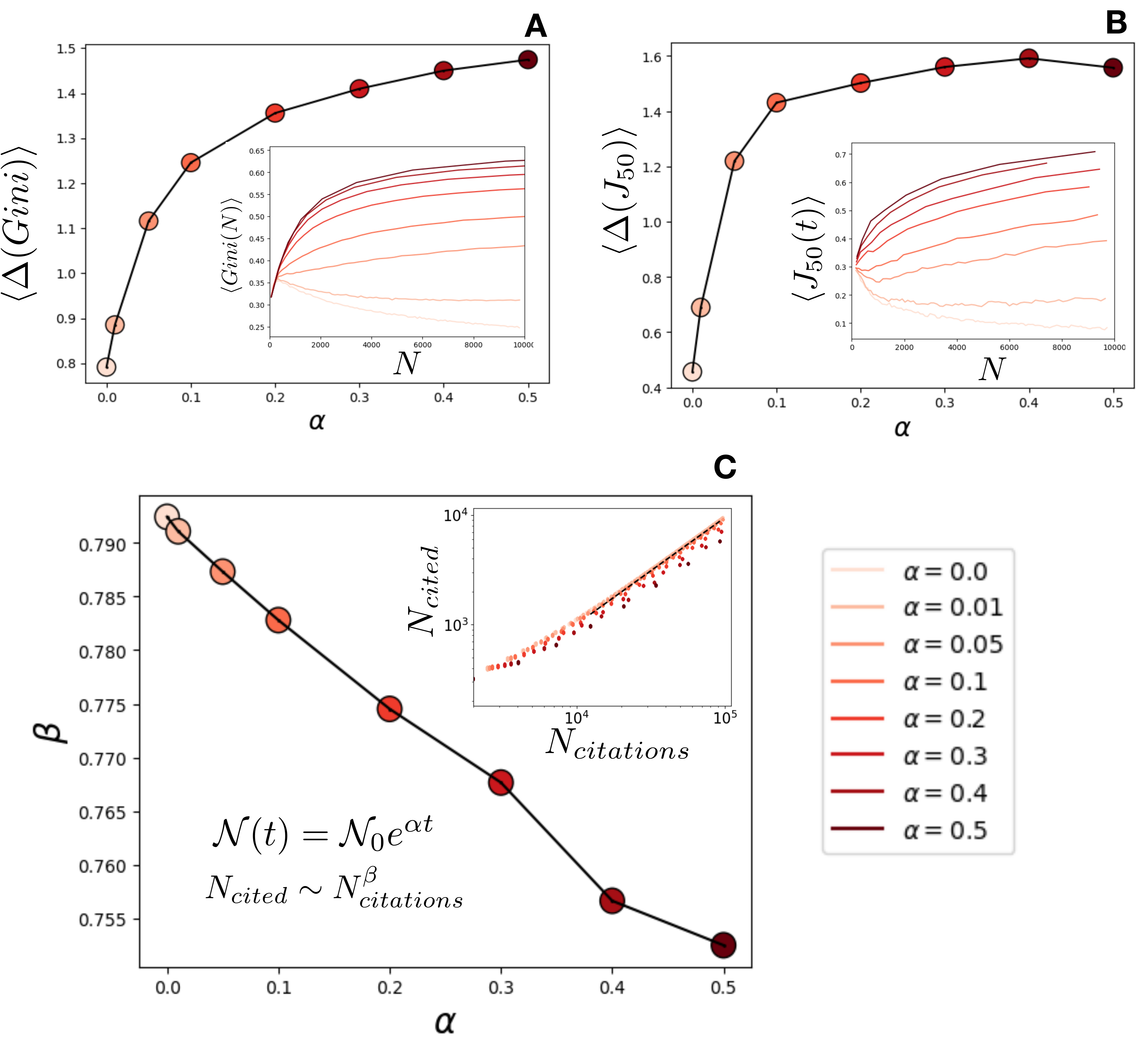}
    \caption{\textbf{(A)} Variation of the Gini index when the population grows from 500 to 5000, as a function of the growth parameter $\alpha$. Inbox: growth of the Gini index as a function of the population size. 
    \textbf{(B)} Variation of the ranked Jaccard index when the population grows from 500 to 5000, respect to the population growth parameter. Inbox: growth of the ranked Jaccard index as a function of the population size. 
    \textbf{(C)} 
    Exponent $\beta$ of the Heaps' law as a function of the parameter $\alpha$.\\
    The results are obtained averaging on 100 replicas of the model for each value of the parameter $\alpha$. 
    \textbf{(C)-inbox}  
    Growth of the number of different publications cited $N_{cited}$ as a function of the total number of citations $N_{citations}$, showing a power-law behaviour typical of the Heaps' law. }
    \label{fig7}
\end{figure}

As we can see from Fig.~\ref{fig7}A and Fig.~\ref{fig7}B, the average value of the Gini index and the ranked Jaccard similarity has a direct dependency with the growth mechanism. We find that t higher values of $\alpha$ correspond to higher values of both the Gini index and the ranked Jaccard similarity. In other words, for the same population size, the concentration of citations among a few publications is much lower when growth occurs slowly (linear if $\alpha=0$) compared to an exponential regime ($\alpha>0$). Similarly, the impact of $\alpha$ on the ranked Jaccard similarity shows that, with stronger exponential growth, elite publications remain increasingly stable at the top each year.

Finally, we can assess the effect of the growth mechanism on the discovery of new literature cited according to our model.
In Fig.~\ref{fig7}C we plot the Heaps' laws of citations for different values of the growth parameter. We see that the value of the Heaps' exponent $\beta$ decreases for larger values of the growth parameter, as shown in Fig.~\ref{fig7}C. This result, consistent with what we observed empirically in Fig.~\ref{fig5}B, shows that higher growing rates correspond to slower exploration processes of the knowledge space. 

\medskip

In its simplicity, this model explains that the presence of a too large number of new papers is compromising the equilibrium between the preferential attachment mechanism (selection in urn $\mathcal{S}$) and the discovery process (selection in urn $\mathcal{U}$). When too many new papers are present in the urn $\mathcal{U}$ each year---a situation that occurs for higher values of $\alpha$---each new paper is cited too infrequently to enter the urn $\mathcal{S}$ with enough visibility to compete with pre-existing star papers. This mechanism amplifies the first-mover advantage and naturally suppresses opportunities for newcomers to compete. 

\section{Conclusions}

Similarly to most other cultural sectors, science has been facing the problem of information overload, due to the exponential increase in the number of scientific publications each year. 
In social media this overload is responsible of several communication troubles, among which the emergence of ``junk news bubbles'', namely ephemeral pieces of content that attract a large share of attention, but that also are quickly forgotten and replaced by newer ones. 
In scientific literature, instead, we do not observe such acceleration of attention cycles. While a smaller and smaller portion of papers monopolizes most citations in each disciplinary arena, this does not produce the quick turnover of popular works observed in social media. On the contrary, popular scientific publications remain so for longer and longer periods of time, creating a growing ``scientific gerontocracy'', where fewer and older elites hold on to the top positions in popularity rankings. As a result, scientific canons stagnate. 

This trend is undoubtedly driven to multiple factors, including the transformation of scientific and editorial practices, the platformization of science, evolving research behaviours in literature exploration, and, more recently, the introduction of Artificial Intelligence (AI) tools. While these elements certainly contribute to the gerontocratization phenomenon, our study shows that the mere fact that science is growing exponentially may itself be a direct cause of this transformation. Indeed, in a world of limited attention, the exponential growth of scientific production makes it increasingly difficult for scientific communities to process and absorb new research.  

To separate the effects of hypergrowth from other processes that could drive the same dynamic, we proposed a simple model allowing to analyze different growth scenarios, while keeping scientific practices constant.
The model shows that a ``gerontocratic'' regime emerges as a direct consequence of the exponential expansion of the system, because of the low probability of selecting new contents with stronger exponential growth rates. 
However, our analysis does not quantify the strength of this causal relationship relative to other endogenous and exogenous confounding factors.

While this model does not fully capture empirical observations or predict future scenarios due to its inherent simplicity, this simplicity was essential to isolate the effect of hypergrowth in the gerontocratization process. 
Future research should build on this framework by incorporating additional  confounding factors.
Among endogenous factors, a key starting point could be the varying intrinsic quality of papers. Accounting for the ``fitness'' of publications would allow an assessment of the impact of ``junk'' papers on the system, as well as the role of predatory journals that publish them. This would also help to evaluate the influence of self-citations and other citation manipulation practices. 
Furthermore, the role of author reputation could be investigated to provide a more complete understanding of the cumulative advantage process. 
Regarding exogenous factors, it is crucial to mention the inherent biases of scientific search engines (from Google Scholar to Elicit), academic platforms (from Academia to ResearchGate)~\cite{ibrahim2024googlescholarmanipulatable} and social media ``opinion leaders"\cite{Haunschild2021}. The impact of these new tools and practices on the paper discovery process remains largely unknown, but their significance could be estimated using a model similar to the one proposed here.
Finally, we must acknowledge the rise of AI-driven research tools, which have the potential to accelerate the entire scientific production chain. 


Overall, this paper, despite some inherent limitations, highlights how science hypergrowth may have triggered a generalized gerontocratization of academic literature as a direct consequence of the information overload in the scientific ecosystem. Understanding this mechanism can help define strategies to reverse or at least mitigate this process and restore a healthy renewal of scientific canons.

\section{Methods}
\subsection{Data Sources}
\label{data}

In this study, we use the OpenAlex (OA) database; this choice was motivated by the open-access availability and the extensive scope of this database \cite{haunschild2024use}. OA offers a comprehensive and unfiltered database, allowing us to include nearly the entire body of scientific publications, including some low-value or ``junk'' papers~\cite{venturini2019fake,lattier2016professors} typically excluded by providers such as Web of Science or Scopus.
Additionally, OA provides a classification system, based on ``concepts'' associated to each paper, that allows us to create fine-grained collections of papers focused on specific disciplines, sub-disciplines, methods or theoretical backgrounds~\cite{priem2022openalexfullyopenindexscholarly}. We hence extracted 14 datasets related to concepts of different nature at the second level of this hierarchical classification.

Finally, to validate our analysis conducted on such OA datasets, we utilized two additional datasets based on the Web Of Science (WOS) database, previously compiled in the context of other projects~\cite{fontaine2023epistemic}.

\subsection{Data Collection}
\label{data_collection}

For our analysis we first identified an initial set of concepts representing a large variety of disciplines having different publishing practices. We then made a first query to the OA database to get the number of works associated with each of these concepts, and randomly selected a subset of concepts with a constrained number ($\approx$ 100\,000) of related papers, so as to ensure that each selected concept represents a bounded sub-field.

To construct the disciplinary datasets based on an OA concept we performed the following steps.
\begin{enumerate}
    \item We first retrieved all papers that contain the selected concept, representing the publications of the related sub-field (disciplinary dataset).
    \item We extended this initial dataset by retrieving all publications citing or cited by the papers obtained in the first step (extended dataset), to get the whole citation history of the selected papers and their relationship with external disciplines.
\end{enumerate}

To construct the disciplinary datasets based on WOS journals we performed the following steps.
\begin{enumerate}
    \item We first identified all scientific journals associated with the selected discipline in WOS and established their correspondence with OA identifiers.
    \item We retrieved, from OpenAlex, all papers published in these journals (disciplinary dataset).
    \item We then collected all papers citing or cited by the papers retrieved during the previous step (extended dataset).
\end{enumerate}

The construction of all datasets have been conducted from September 2023 to January 2024. 
In these datasets, we only considered papers published from 1980 to 2019. 
Additionally, notice that an initial filtering was made using the OpenAlex API, in order to only consider published articles, i.e., including a DOI, and excluding retracted papers and documents including no references. 
Finally, notice that when we refer to a discipline in our analysis, we specifically indicate the papers in the disciplinary dataset. Moreover, the citations that a paper receives are counted on the extended datasets, including the papers that do not contain the selected concept (or journals).

A summary of the selected OA concepts or WOS fields of study and the size of their related disciplinary and extended dataset dataset is provided in Table~\ref{table1}.

\begin{table}[ht]
\centering
\begin{tabular}{lrr}
\toprule
Discipline & Dataset size & Extended dataset size\\
\midrule
Neuroscience (WOS) & 1,578,556 & 11,888,972 \\
Sociology (WOS) & 949,605 & 6,015,791 \\
Cognitive science & 148,027 & 2,661,581 \\
Computer architecture & 85,470 & 507,037 \\
Food security & 113,283 & 609,721 \\
Labor economics & 112,075 & 1,234,409 \\
Molecular physics & 160,524 & 2,318,947 \\
Cryptography & 80,314 & 318,990 \\
Architecture & 126,215 & 1,516,638 \\
Biochemical engineering & 110,452 & 3,695,296 \\
Speech recognition & 207,264 & 1,632,350 \\
Water resource management & 75,050 & 818,540 \\
Computational physics & 177,372 & 1,939,356 \\
Criminology & 178,532 & 1,533,333 \\
Graph & 177,914 & 1,666,413 \\
Grammar & 24,367 & 301,240 \\
\bottomrule
\end{tabular}
\caption{Size information of the disciplinary and extended datasets related to the selected OA concepts and WOS fields of study.}
\label{table1}
\end{table}

\subsection{Model setup}
\label{methods-model}
The model used in section \ref{Q2} simulates the formation of citation networks in an expanding publication system. This expansion follows the exponential growth pattern observed in the data ($\mathcal{N}(t)=\mathcal{N}_0e^{\alpha t}$).
Each new paper entering the system at time $t$ selects a set of $N_{ref}$ references, making a stochastic choice between already established papers and discovering new scientific works. The parameter $p$ controls the probability of choosing between these two options and is calibrated based on empirical observations. 
\begin{figure}[tb]
    \centering
    \includegraphics[width=8cm]{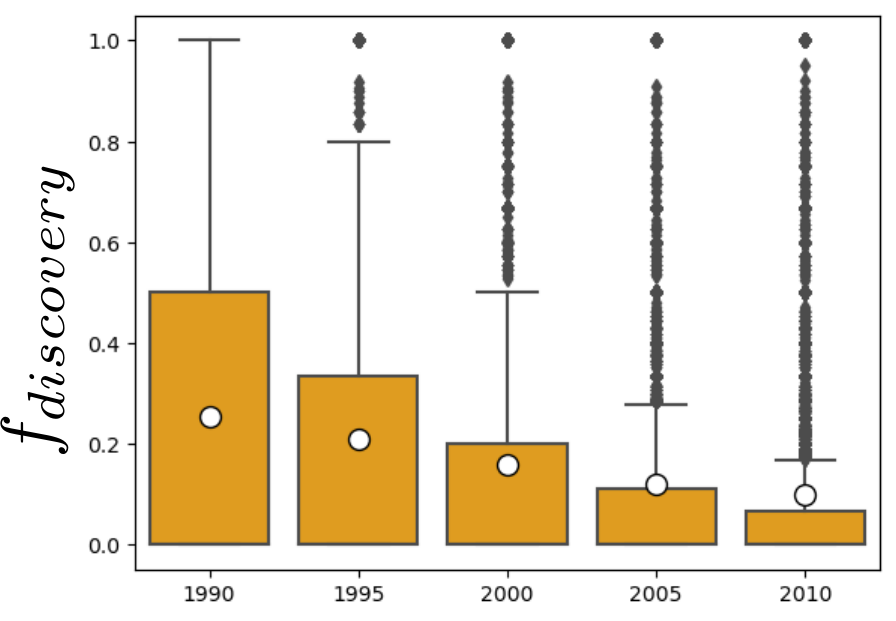}
    \caption{Fraction of previously uncited papers in article reference lists over time, considering all disciplines together.}
    \label{fig8}
\end{figure}
In particular, in order to determine the value of the parameter $p$ in our model, we examined the fraction of citations $f_{discovery}$ directed to previously uncited papers. For this analysis, each reference list was filtered to include only papers indexed in the original dataset (without any filtering). Such choice ensures a complete citation history for each paper in our datasets. 


Since the results are consistent across disciplines, we aggregate all disciplines together, and show the evolution of $f_{discovery}$ in Fig.~\ref{fig8}. 
Here, we notice that the parameter $p=1-f_{discovery}$ is slightly increasing over time. 
This finding points to a shift in individual author choices, increasingly reinforcing the Matthew effect. This trend may also be influenced by the recommendation algorithms of academic search engines (such as Google Scholar), an effect that warrants further investigation in future studies.
However, to isolate the effects of system growth from those potentially caused by fluctuations of this parameter, we fixed $p = 0.8$ in all our simulations, representing the ``best scenario'' observed in the data. 

Moreover, the parameter $N_{ref}$ is fixed to 10 in the simulations, although this value does not impact the model’s outcomes.
Furthermore, at the beginning of a simulation, the urn $\mathcal{S}$ is initialized with a set of 200 papers---on the order of the number of cited papers in 1980 in the data---, with an initial citation count uniformly distributed between 1 and 3. 
The urn $\mathcal{U}$ initially contains 100 papers, i.e., 50\% of the papers initially in $\mathcal{S}$, reflecting that, on average, 50\% of papers remains uncited. The paper ages are randomly assigned between 1 and 2 years, and the number of new papers appearing in the first iteration of the simulation is set to half the combined size of the two urns, and then continues growing according to the rule $\mathcal{N}(t)=\mathcal{N}_0e^{\alpha t}$.
The process is iterated until the total number of papers reaches a final value of 50\,000. For each value of the parameter $\alpha$, from $\alpha = 0$ to $\alpha = 0.5$, we perform 100 replicas of the simulation.

\backmatter





\bmhead{Acknowledgements}
The authors acknowledge the support of the French Agence Nationale de la Recherche (ANR), under grant  ANR-21-CE38-0020 (project ScientIA). 
The authors would like to thank A. Maddi for the useful discussions.

\bibliography{sample}

\end{document}